\documentclass[aps,prb,showpacs,floatfix,twocolumn,byrevtex,superscriptaddress]{revtex4-1}

\usepackage[english]{babel}
\usepackage{amsmath}
\usepackage{graphicx}
\usepackage{float}
\usepackage{bm}
\usepackage{multirow}
\usepackage{dcolumn}
\usepackage{hyperref}

\newcommand{\icm}{\ensuremath{~\textrm{cm}^{-1}}}
\newcommand{\BKFAx}{Ba$_{1-x}$K$_{x}$Fe$_{2}$As$_{2}$}
\newcommand{\BFAPx}{BaFe$_{2}$(As$_{1-x}$P$_{x}$)$_{2}$}
\newcommand{\BFCAx}{Ba(Fe$_{1-x}$Co$_{x}$)$_{2}$As$_{2}$}

\begin{document}

\title{\boldmath Optical observation of spin-density-wave fluctuations in Ba122 iron-based superconductors \unboldmath}
\author{B. Xu}
\affiliation{Center for High Pressure Science and Technology Advanced Research, Beijing 100094, China}
\affiliation{Beijing National Laboratory for Condensed Matter Physics, Institute of Physics, Chinese Academy of Sciences, Beijing 100190, China}
\affiliation{LPEM, ESPCI Paris, PSL Research University, CNRS, 10 rue Vauquelin, F-75231 Paris Cedex 5, France}

\author{Y. M. Dai}
\email[]{ymdai@lanl.gov}
\affiliation{Beijing National Laboratory for Condensed Matter Physics, Institute of Physics, Chinese Academy of Sciences, Beijing 100190, China}
\affiliation{LPEM, ESPCI Paris, PSL Research University, CNRS, 10 rue Vauquelin, F-75231 Paris Cedex 5, France}
\affiliation{Universit\'e Pierre et Marie Curie, Sorbonne Universit\'es, F-75005 Paris Cedex 5, France}

\author{H. Xiao}
\affiliation{Center for High Pressure Science and Technology Advanced Research, Beijing 100094, China}

\author{B. Shen}
\affiliation{National Laboratory of Solid State Microstructures and Department of Physics, Nanjing University, Nanjing 210093, China}

\author{Z. R. Ye}
\affiliation{State Key Laboratory of Surface Physics, Department of Physics, and Advanced Materials Laboratory, Fudan University, Shanghai 200433, China}

\author{A. Forget}
\author{D. Colson}
\affiliation{IRAMIS, SPEC, CEA, 91191 Gif sur Yvette, France}

\author{D. L. Feng}
\affiliation{State Key Laboratory of Surface Physics, Department of Physics, and Advanced Materials Laboratory, Fudan University, Shanghai 200433, China}

\author{H. H. Wen}
\affiliation{National Laboratory of Solid State Microstructures and Department of Physics, Nanjing University, Nanjing 210093, China}

\author{X. G. Qiu}
\affiliation{Beijing National Laboratory for Condensed Matter Physics, Institute of Physics, Chinese Academy of Sciences, Beijing 100190, China}

\author{R. P. S. M. Lobo}
\email[]{lobo@espci.fr}
\affiliation{LPEM, ESPCI Paris, PSL Research University, CNRS, 10 rue Vauquelin, F-75231 Paris Cedex 5, France}
\affiliation{Universit\'e Pierre et Marie Curie, Sorbonne Universit\'es, F-75005 Paris Cedex 5, France}

\date{\today}

%

\begin{abstract}
In iron-based superconductors, a spin-density-wave (SDW) magnetic order is suppressed with doping and unconventional superconductivity appears in close proximity to the SDW instability. The optical response of the SDW order shows clear gap features: substantial suppression in the low-frequency optical conductivity, alongside a spectral weight transfer from low to high frequencies. Here, we study the detailed temperature dependence of the optical response in three different series of the Ba122 system [\BKFAx, \BFCAx\ and \BFAPx]. Intriguingly, we found that the suppression of the low-frequency optical conductivity and spectral weight transfer appear at a temperature $T^{\ast}$ much higher than the SDW transition temperature $T_{SDW}$. Since this behavior has the same optical feature and energy scale as the SDW order, we attribute it to SDW fluctuations. Furthermore, $T^{\ast}$ is suppressed with doping, closely following the doping dependence of the nematic fluctuations detected by other techniques. These results suggest that the magnetic and nematic orders have an intimate relationship, in favor of the magnetic-fluctuation-driven nematicity scenario in iron-based superconductors.
\end{abstract}



\maketitle

%
Iron-based superconductors (FeSCs) feature complex phase diagrams with multiple phase transitions, including unconventional superconductivity, magnetic and nematic phases. Unconventional superconductivity always appears in the vicinity of other competing phases.~\cite{Paglione2010} In FeSCs, the main competing phases are the SDW magnetic and nematic orders. By applying carrier doping, isovalent substitution, or pressure, the SDW and nematic orders are suppressed and unconventional superconductivity emerges.~\cite{Paglione2010,Basov2011}

In FeSCs, the competing phases usually consist of various symmetry-breaking orders, which interact with superconductivity in a complicated way.~\cite{Paglione2010,Basov2011} The SDW magnetic order is of the stripe type (spins aligned ferromagnetically in one direction and antiferromagnetically in the other).~\cite{Inosov2009,PCDai2012} Such an order breaks the $O(3)$ spin-rotational symmetry.~\cite{Fernandes2014} The nematic order breaks the $C_{4}$ symmetry, which is characterized by inequivalent $a$ and $b$ lattice parameters and anisotropic electronic properties.~\cite{Chu2010,Yi2011,Tanatar2010,Kasahara2012,Fernandes2012,Nakajima2011a,Nakajima2012,Dusza2011a} The nematic order often occurs either simultaneously or slightly above the SDW transition, making these two orders track each other closely. At present, there are two scenarios for the development of nematic order and its relation to SDW magnetic order. One scenario is that nematic order is driven by orbital ordering of the iron $3d$ electrons. The orbital fluctuations trigger the magnetic transition and induce the striped SDW order.~\cite{Kruger2009,Lv2009,Chen2010} Another scenario is that the nematic order is driven by magnetic fluctuations, which produce a spin-nematic phase associated with the fact that the striped SDW order can be along either $a$ or $b$ axis.~\cite{Eremin2010,Xu2008,Fernandes2012PRB,Avci2014} However, it is not easy to assess the relative importance of orbital fluctuations and spin fluctuations from measurable quantities, making the origin of nematicity remain a controversial issue.~\cite{Fernandes2014} It is known that the superconducting order parameter due to spin fluctuations has $s_{\pm}$ symmetry,~\cite{Mazin2008,Kuroki2008} whereas the one due to orbital fluctuations has $s_{++}$ symmetry.~\cite{Kontani2010,Kontani2011} Therefore, to understand the unconventional superconductivity in FeSCs, it is important to investigate these competing orders, as well as corresponding fluctuations, since the unconventional superconductivity may be driven by these fluctuations.~\cite{Fernandes2014}

\begin{figure*}[tb]
\includegraphics[width=\textwidth]{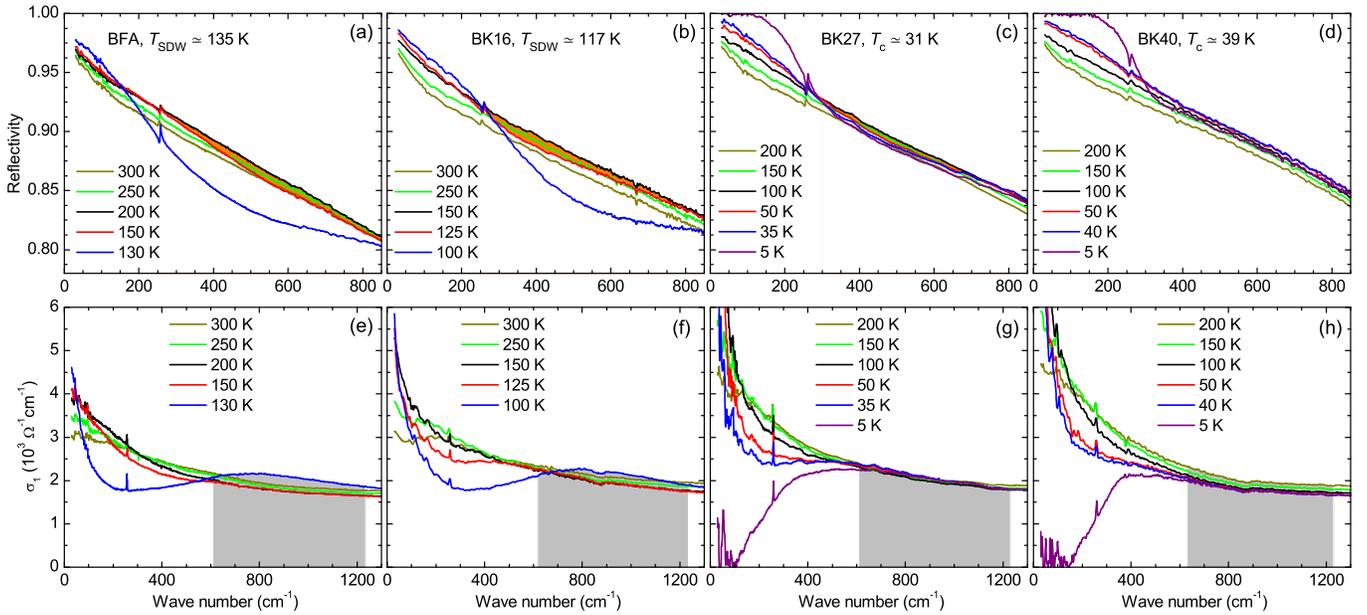}
\caption{ (Color online)  (a--d) show the far-infrared reflectivity at different temperatures for BFA, BK16, BK27 and BK40, respectively. (e--h) display the corresponding optical conductivity curves obtained by the Kramers--Kronig analysis. The orange filled area indicates the signature of the anomalous behavior above $T_{SDW}$. The grey area indicates the spectral weight transfer region due to the opening of the SDW gap.}
\label{Fig1}
\end{figure*}
Optical spectroscopy probes the charge dynamics in solid materials, providing important information about phase transitions. The optical response of the SDW and SC transitions in FeSCs has been studied by many previous works,~\cite{Hu2008,Li2008,Lobo2010,Dai2012PRB,Dai2013a,Dai2016} where clear gap features have been identified. However, optical investigations into these materials with detailed temperature and doping dependence across the SDW or nematic transitions are still lacking, since a large amount of measurements are required. Here, we fill this gap by presenting a systematic optical study on a series of high quality K-, Co-, and P-doped Ba122 single crystals. We observed strong SDW fluctuations above the SDW transition temperature. The onset SDW fluctuations temperature $T^{\ast}$ is defined by tracking the temperature evolution of the reflectivity and spectral weight. By plotting $T^{\ast}$ in the phase diagrams, we find that it is in good agreement with the doping dependence of nematic fluctuations reported by previous works,~\cite{Blomberg2013,Kasahara2012} suggesting a close connection between the magnetic and nematic orders.

%
In this work, we study three series of Ba122 FeSCs, including \BKFAx\ (Ref.~\onlinecite{Shen2011}), \BFCAx\ (Ref.~\onlinecite{Rullier-Albenque2009}), and \BFAPx\ (Ref.~\onlinecite{Ye2012}). For each series, high-quality single crystals of various dopings were synthesized with the methods described in the cited references. The samples are named by their dopant percentages throughout the paper. For example, the $x$ = 0, 0.08, 0.16, 0.20, 0.23, 0.27, and 0.40 samples of \BKFAx\ are named BFA, BK8, BK16, BK20, BK23, BK27, and BK40, respectively. In all three series, as shown by the transport properties in Appendix~\ref{apppendixA}, the undoped and underdoped compounds show a magnetic transition at $T_{SDW}$, while the superconducting compounds undergo a superconducting transition at $T_c$.

\begin{figure*}[tb]
\includegraphics[width=\textwidth]{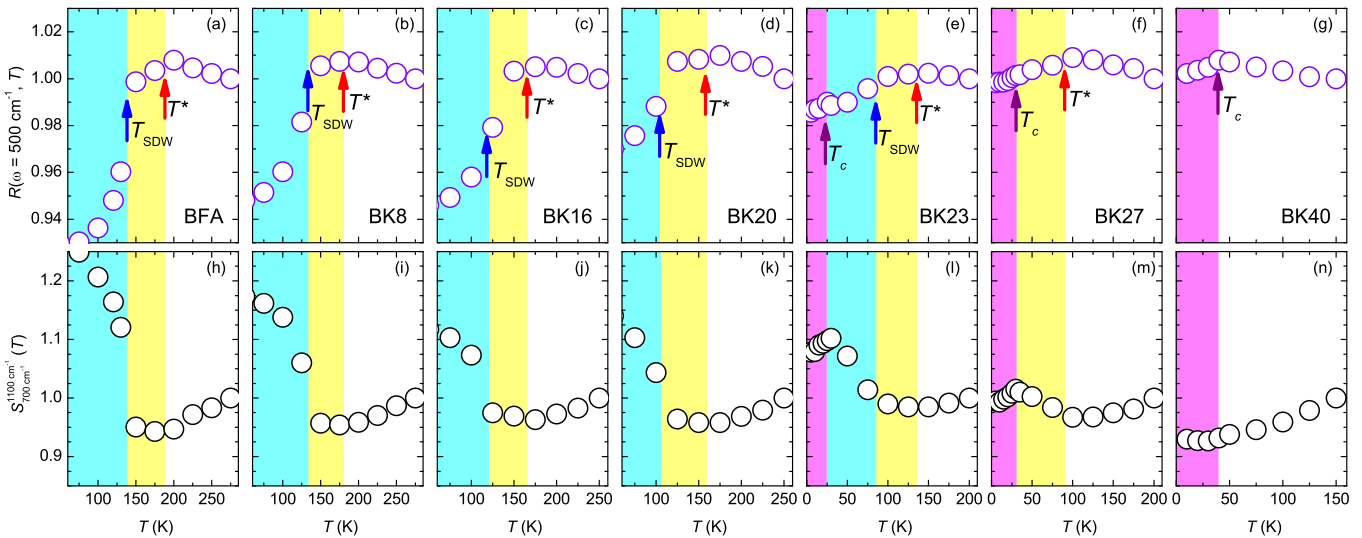}
\caption{ (Color online) Temperature dependence of $R(\omega = 500\icm)$ (a--g) and $S_{700\icm}^{1\,100\icm}$ (h--n) in \BKFAx. All values are normalized to the highest temperature given in the panels for the best comparison. The red, blue and purple arrows indicate the onset temperature of the fluctuations $T^{\ast}$, SDW transition temperature $T_{SDW}$ and superconducting transition temperature $T_c$, respectively.}
\label{Fig2}
\end{figure*}
The \emph{ab}-plane reflectivity, $R(\omega)$, was measured at a near-normal angle of incidence on Bruker IFS66v and IFS113v spectrometers. An \emph{in situ} gold overfilling technique~\cite{Homes1993} was used to obtain the absolute reflectivity of the samples. Data from 30 to $15\,000\icm$ were collected at different temperatures on freshly cleaved surface for each sample, and then we extended the reflectivity to $40\,000\icm$ at room temperature with an AvaSpec-2048 $\times$ 14 optical fiber spectrometer. The optical conductivity, $\sigma_{1}(\omega)$, was obtained by Kramers-Kronig analysis of the reflectivity. At low frequencies, we employed a Hagen-Rubens ($R = 1 - A\sqrt{\omega}$) or a superconducting ($R = 1 - A\omega^4$) extrapolation. Above $15\,000\icm$, for all temperatures, we utilized the room temperature data, followed by a constant reflectivity up to 12.5~eV, and terminated by a free-electron ($\omega^{-4}$) response.

%
\begin{figure}[tb]
\includegraphics[width=0.85\columnwidth]{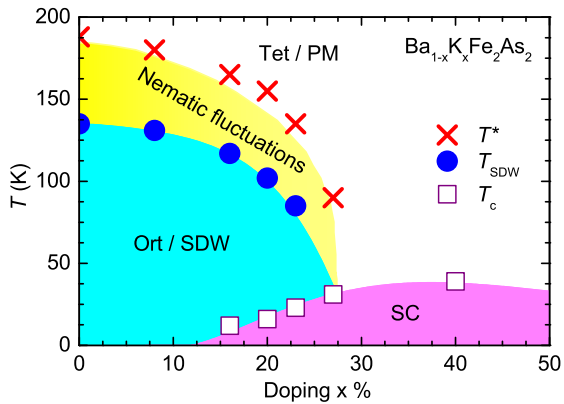}
\caption{(Color online) Phase diagram of \BKFAx. The crosses denote the onset temperature $T^{\ast}$ of the fluctuations. The solid circles mark the SDW transition temperature $T_{SDW}$. The open squares represent the superconducting transition temperature $T_{c}$. The background of the phase diagram was extracted from Ref.~\onlinecite{Blomberg2013}. The superconducting, orthorhombic SDW, and tetragonal paramagnetic phases are abbreviated as SC, Ort/SDW, and Tet/PM, respectively.}
\label{Fig3}
\end{figure}
We start with our data analysis in the hole-doped compounds \BKFAx. In this system, the optical response is expected to exhibit distinct changes at each phase transition. Figure~\ref{Fig1} shows $R(\omega)$ in panels (a--d) and $\sigma_1(\omega)$ in panels (e--h) for \BKFAx\ at different temperatures. Generally, the normal state ($T > T_{SDW}, T_{c}$) optical spectra exhibit typical metallic response, realized by the relatively high $R(\omega)$ and the Drude-like behavior in $\sigma_1(\omega)$. As the temperature is lower, $R(\omega)$ increases continuously, but is suppressed in the frequency range of 200--900\icm\ for $T < T_{SDW}$ in BFA and BK16. Meanwhile, the corresponding $\sigma_{1}(\omega)$ is also severely suppressed and most of spectral weight below $\sim$~600\icm\ is transferred to higher-energy region forming a peak around 900\icm. This spectral evolution manifests the behavior of the SDW gap on the Fermi surface.~\cite{Hu2008,Dai2012PRB} Upon entering the superconducting state, the low-frequency $R(\omega)$ develops a sharp edge and rises to a flat 100\% value, as shown by the 5~K data for BK27 and BK40. This is a signature of the superconducting gap. In BK27 and BK40, $\sigma_{1}(\omega)$ at 5~K drops to zero at low frequencies, indicating a fully open superconducting gap.~\cite{Li2008,Dai2013a}

In addition to these distinct spectral signatures associated with the SDW and superconducting gaps, we observe  an anomalous behavior. As indicated by the orange filled area in Fig.~\ref{Fig1}(a--c), above $T_{SDW}$, a small suppression of $R(\omega)$ sets in between $\sim$~300 and 800\icm, which falls into the energy scale of the SDW gap. This behavior can also be recognized on $\sigma_{1}(\omega)$. For instance, in BFA [Fig.~\ref{Fig1}(e)], $\sigma_{1}(\omega)$ decreases continuously from 300 to 200~K in the grey filled region ($\sim$~600--1\,200\icm), owing to the narrowing of the Drude peak. However, we note that at 150~K the value of $\sigma_{1}(\omega)$ becomes slightly higher than that at 200~K. This is explicit evidence of the spectral weight transfer from low to high frequencies above $T_{SDW}$. With more doping, this behavior becomes weaker in BK27, and finally disappears in the optimally-doped compound BK40 above the SC transition.

\begin{figure*}[tb]
\includegraphics[width=\textwidth]{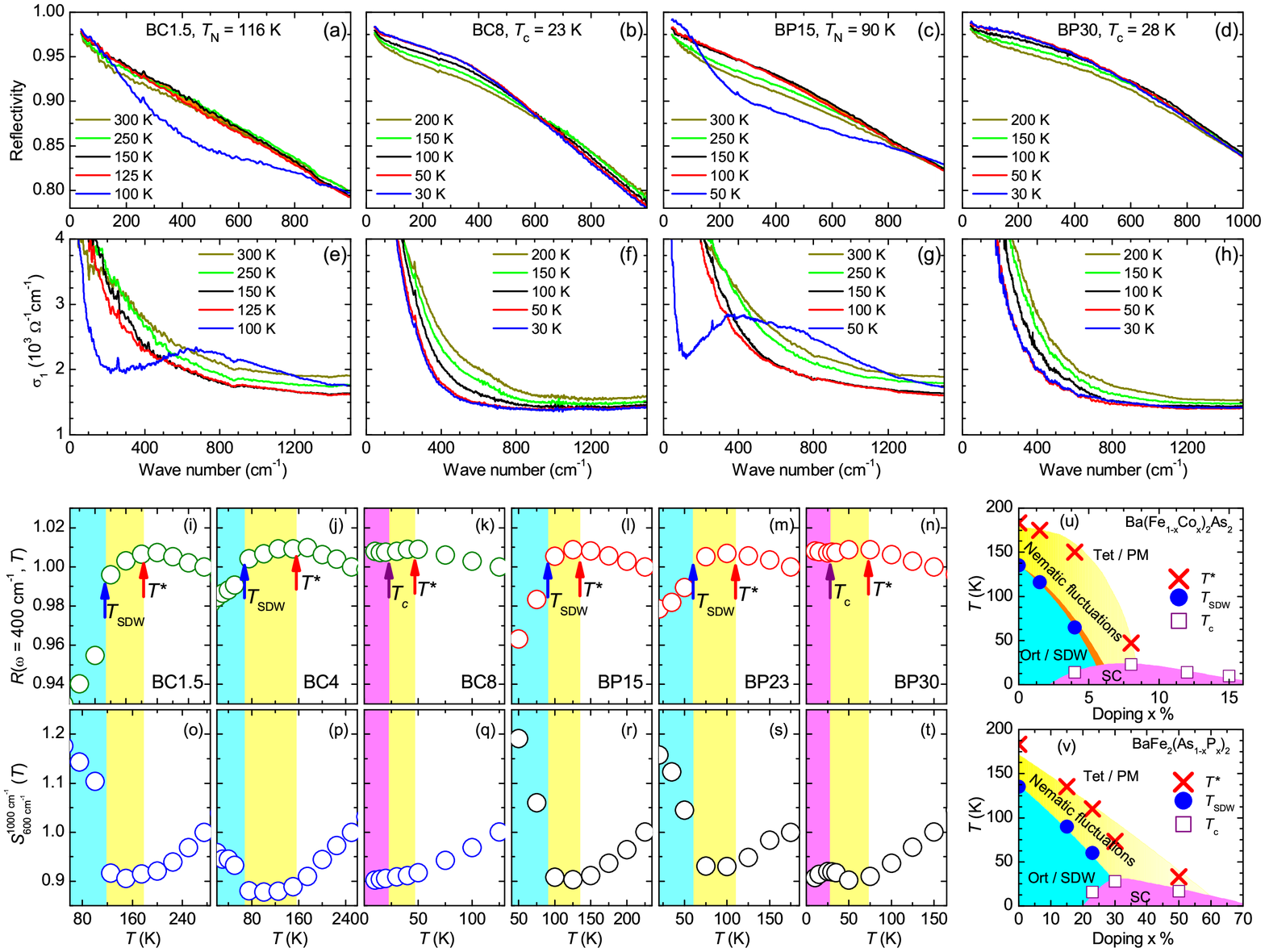}
\caption{ (Color online) Far-infrared reflectivity (a--d) and optical conductivity (e--h) for BC1.5, BC8, BP15 and BP30, respectively. Temperature dependence of $R(\omega = 400\icm)$ (i--n) and $S_{600\icm}^{1\,000\icm}$ (o--t) for \BFCAx\ and \BFAPx. All values are normalized to the highest temperature given in the panels for the best comparison. The red, blue and purple arrows indicate $T^{\ast}$, $T_{SDW}$ and $T_c$, respectively. (u) and (v) depict the phase diagrams of \BFCAx\ and \BFAPx, respectively. The background of the phase diagrams were extracted from Refs.~\onlinecite{Blomberg2013,Kasahara2012}.}
\label{Fig4}
\end{figure*}
In the following, we determine the onset temperature of the anomalous behavior observed above $T_{SDW}$. One convenient way is to track the temperature dependence of $R(\omega)$ at a certain frequency in the suppressed region. Here, we take the value of $R(\omega)$ at 500\icm, $R(\omega = 500\icm)$, and plot it as a function of temperature for each doping [Fig.~\ref{Fig2}(a--g)]. In the parent compound BFA, as shown in Fig.~\ref{Fig2}(a), $R(\omega = 500\icm)$ increases as the temperature is reduced, which is expected for a metallic behavior. In sharp contrast, $R(\omega = 500\icm)$ begins to decrease at a temperature $T^{{\ast}}$ (indicated by the red arrow at $\simeq$ 185~K). Below $T_{SDW}$ (indicated by the blue arrow), $R(\omega = 500\icm)$ exhibits a more dramatic drop due to the opening of the SDW gap. In other samples, similar behavior is revealed, allowing the determination of the onset temperature $T^{\ast}$ for each doping. However, it is worth pointing out that $R(\omega = 500\icm)$ does not show any anomaly above $T_c$ in BK40, implying that this behavior is absent.

Anomalies at $T^{\ast}$ can also be verified by a spectral weight analysis. The spectral weight is defined as $S_{\omega_{a}}^{\omega_{b}} = \int_{\omega_{a}}^{\omega_{b}}\sigma_{1}(\omega)d\omega$, where $\omega_{a}$ and $\omega_{b}$ are the lower and upper cutoff frequency, respectively. Figs.~\ref{Fig2}(h--n) depict the spectral weight between 700--1\,100\icm, $S_{700\icm}^{1\,100\icm}$, for each doping. $S_{700\icm}^{1\,100\icm}$ decreases continuously upon cooling due to the narrowing of the Drude peak, but starts to increase at $T^{\ast}$ signaling the onset of the anomalous spectral weight transfer from low to high frequencies. Below $T_{SDW}$, the increase of $S_{700\icm}^{1\,100\icm}$ becomes more substantial, which is caused the opening of the SDW gap. In Fig.~\ref{Fig3}, we summarize the doping evolution of $T^{\ast}$ (red crosses) in the phase diagram of \BKFAx: with increasing K concentration, $T^{\ast}$ is suppressed and terminates on the border of the SDW phase.

In order to confirm whether these observations are common in all Ba122 compounds, it is instructive to look into the electron-doped \BFCAx\ and isovalent-doped \BFAPx. Figure ~\ref{Fig4} manifests that these two series show identical optical response above the SDW transition. $T^{\ast}$ is also extracted from the temperature dependence of $R(\omega)$ and spectral weight. The phase diagrams in Fig.~\ref{Fig4}(u) and Fig.~\ref{Fig4}(v) summarized the doping evolution of $T^{\ast}$ in \BFCAx\ and \BFAPx, respectively. In both series, $T^{\ast}$ is suppressed with increasing doping, following the trend of the SDW order. Interestingly, $T^{\ast}$ survives up to the optimal doping in \BFCAx, and the overdoping in \BFAPx.

Based on the above optical results in all three series of the Ba122 system, we summarize our new findings: (i) the onset temperature ($T^{\ast}$) of the low-frequency $\sigma_1(\omega)$ suppression and spectral weight transfer is above the SDW transition; (ii) the optical response and energy scale of the above anomalous behavior are the same as the SDW gap; (iii) the doping evolution of $T^{\ast}$ is similar to the SDW order; (iv) $T^{\ast}$ disappears in the underdoped region for \BKFAx, while it survives up to the optimally-doped and overdoped region in \BFCAx\ and \BFAPx, respectively. These facts suggest that the phenomenon we observed is likely to arise from SDW fluctuations, and these fluctuations span different regions of the phase diagram for different doping types.~\cite{Wilson2010,Ning2010,Nakai2010b,Li2011,Moon2014}

Having attributed the observed anomalous behavior to SDW fluctuations, we now discuss its relation to nematicity in these compounds. In FeSCs, the electronic nematicity is characterized by anisotropic properties along the $a$ and $b$ axis.~\cite{Chu2010,Blomberg2013,Kasahara2012,Yi2011,Tanatar2010,Fernandes2012,Blomberg2013} In the SDW state, anisotropic energy gap and density of states have been revealed by previous theoretical work.~\cite{Yin2011NP} The $yz$ orbital has fewer electronic states at the Fermi level than the $xz$ orbital. Therefore, the optical response of the SDW order is affected by these anisotropic electronic states. In the detwinned Co-doped Ba122 FeSCs, anisotropic optical properties have been experimentally observed below and slightly above the SDW transition temperature.~\cite{Nakajima2011a,Dusza2011a,Nakajima2012} Although optical measurement on twinned crystals is not sensitive to the nematic order or fluctuations, we found that, as shown by the phase diagrams in Figs.~\ref{Fig3}, \ref{Fig4}(u) and \ref{Fig4}(v), $T^{\ast}$ follows the reported doping dependence of nematic fluctuations~\cite{Blomberg2013,Kasahara2012} (yellow areas) quite well in all three series. This implies that SDW fluctuations and the nematicity are closed tied to each other. More interestingly, in \BKFAx, where SDW fluctuations vanish in the underdoped region, nematic fluctuations disappear at approximately the same doping, while in \BFAPx, where SDW fluctuations exist up to the overdoped region, nematic fluctuations persist to the overdoped regime. These observations indicate that the SDW and nematic fluctuations are governed by the same physics in all three series, which is in favor of the spin-fluctuation-driven nematicity scenario.

The distinct doping dependence of $T^{\ast}$ in \BKFAx, \BFCAx\ and \BFAPx\ may result from the complicated interaction between magnetism, nematicity and superconductivity, which depends on doping types. Recently, in the hole-doped Ba122 FeSCs, a $C_{4}$-symmetric magnetic phase has been revealed by the neutron scattering and thermodynamic experiments.~\cite{Avci2014,Boehmer2015} This $C_{4}$ magnetic phase exists at the boundary between the superconductivity and SDW magnetic order in the phase diagram, reflecting the strong competition between the magnetism, nematicity and superconductivity. In \BKFAx, the termination of the nematic fluctuations in the underdoped regime is likely a direct result of the existence of this $C_{4}$ state. $T^{\ast}$ also disappears at the same doping where the nematic order vanishes, suggesting the intimate relation between nematicity and SDW magnetism.

%
To summarize, we performed systematic optical studies on three series of Ba122 single crystals, including \BKFAx, \BFCAx\ and \BFAPx. In the underdoped regime, optical evidence for strong SDW fluctuations has been revealed. As $x$ approaches the optimal doping, this behavior disappears in the K-doped compounds and becomes weaker in the Co- and P-doped compounds. By examining the doping evolution of the onset temperature of the SDW fluctuations, we found it agrees very well with the doping dependence of nematic fluctuations measured by other techniques. Our observations point to an intimate relationship between magnetic fluctuations and nematicity, supporting the spin-fluctuation-driven namaticity scenario.~\cite{Fernandes2014,Avci2014}

%
This work was supported by NSFC, Grant No. U1530402. Work at IOP CAS was supported by MOST (973 Projects No. 2015CB921303, 2015CB921102, 2012CB921302, and 2012CB821403), and NSFC (Grants No. 91121004, 91421304, and 11374345).

%
\appendix

\section{Transport properties}
\label{apppendixA}
\begin{figure}[tbh]
\includegraphics[width=0.95\columnwidth]{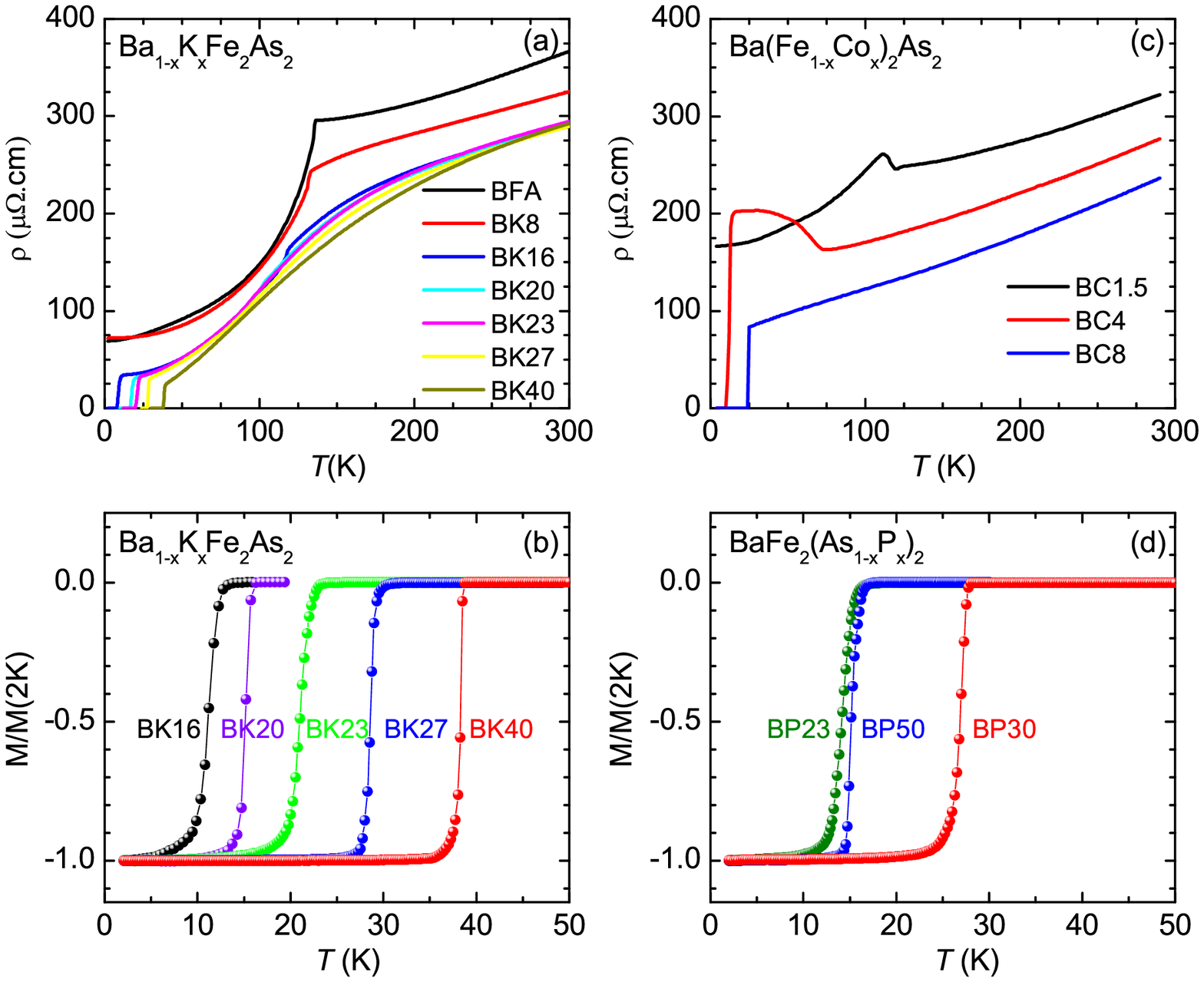}
\caption{(Color online) Temperature dependence of (a) the in-plane resistivity and (b) the magnetic susceptibility in \BKFAx. (c) Temperature dependence of the in-plane resistivity in \BFCAx. (d) Temperature dependence of the magnetic susceptibility in \BFAPx.}
\label{Fig5}
\end{figure}

Fig.~\ref{Fig5}(a) shows temperature dependence of the in-plane resistivity $\rho(T)$ for \BKFAx. Resistivity of the undoped and underdoped compounds shows a sharp drop at $T_{SDW}$ and a metallic behavior below $T_{SDW}$. For the superconducting compounds, such as BK40, $\rho(T)$ shows a sharp SC transition. Fig.~\ref{Fig5}(b) shows the magnetic susceptibility as a function of temperature, measured in a 10~Oe magnetic field. $T_c$ can be defined by the onset of the zero-field-cooling diamagnetic susceptibility. Similarly, Fig.~\ref{Fig5}(c) and Fig.~\ref{Fig5}(d) display the transport properties in \BFCAx\ and \BFAPx, respectively.
%

%


\begin{thebibliography}{42}
\expandafter\ifx\csname natexlab\endcsname\relax\def\natexlab#1{#1}\fi
\expandafter\ifx\csname bibnamefont\endcsname\relax
  \def\bibnamefont#1{#1}\fi
\expandafter\ifx\csname bibfnamefont\endcsname\relax
  \def\bibfnamefont#1{#1}\fi
\expandafter\ifx\csname citenamefont\endcsname\relax
  \def\citenamefont#1{#1}\fi
\expandafter\ifx\csname url\endcsname\relax
  \def\url#1{\texttt{#1}}\fi
\expandafter\ifx\csname urlprefix\endcsname\relax\def\urlprefix{URL }\fi
\providecommand{\bibinfo}[2]{#2}
\providecommand{\eprint}[2][]{\url{#2}}

\bibitem[{\citenamefont{Paglione and Greene}(2010)}]{Paglione2010}
\bibinfo{author}{\bibfnamefont{J.}~\bibnamefont{Paglione}} \bibnamefont{and}
  \bibinfo{author}{\bibfnamefont{R.~L.} \bibnamefont{Greene}},
  \bibinfo{journal}{Nature Physics} \textbf{\bibinfo{volume}{6}},
  \bibinfo{pages}{645} (\bibinfo{year}{2010}).

\bibitem[{\citenamefont{Basov and Chubukov}(2011)}]{Basov2011}
\bibinfo{author}{\bibfnamefont{D.~N.} \bibnamefont{Basov}} \bibnamefont{and}
  \bibinfo{author}{\bibfnamefont{A.~V.} \bibnamefont{Chubukov}},
  \bibinfo{journal}{Nat. Phys.} \textbf{\bibinfo{volume}{7}},
  \bibinfo{pages}{272} (\bibinfo{year}{2011}).

\bibitem[{\citenamefont{Inosov et~al.}(2009)\citenamefont{Inosov, Park,
  Bourges, Sun, Sidis, Schneidewind, Hradil, Haug, Lin, Keimer
  et~al.}}]{Inosov2009}
\bibinfo{author}{\bibfnamefont{D.~S.} \bibnamefont{Inosov}},
  \bibinfo{author}{\bibfnamefont{J.~T.} \bibnamefont{Park}},
  \bibinfo{author}{\bibfnamefont{P.}~\bibnamefont{Bourges}},
  \bibinfo{author}{\bibfnamefont{D.~L.} \bibnamefont{Sun}},
  \bibinfo{author}{\bibfnamefont{Y.}~\bibnamefont{Sidis}},
  \bibinfo{author}{\bibfnamefont{a.}~\bibnamefont{Schneidewind}},
  \bibinfo{author}{\bibfnamefont{K.}~\bibnamefont{Hradil}},
  \bibinfo{author}{\bibfnamefont{D.}~\bibnamefont{Haug}},
  \bibinfo{author}{\bibfnamefont{C.~T.} \bibnamefont{Lin}},
  \bibinfo{author}{\bibfnamefont{B.}~\bibnamefont{Keimer}},
  \bibnamefont{et~al.}, \bibinfo{journal}{Nat. Phys.}
  \textbf{\bibinfo{volume}{6}}, \bibinfo{pages}{178} (\bibinfo{year}{2009}).

\bibitem[{\citenamefont{Dai et~al.}(2012{\natexlab{a}})\citenamefont{Dai, Hu,
  and Dagotto}}]{PCDai2012}
\bibinfo{author}{\bibfnamefont{P.}~\bibnamefont{Dai}},
  \bibinfo{author}{\bibfnamefont{J.}~\bibnamefont{Hu}}, \bibnamefont{and}
  \bibinfo{author}{\bibfnamefont{E.}~\bibnamefont{Dagotto}},
  \bibinfo{journal}{Nat. Phys.} \textbf{\bibinfo{volume}{8}},
  \bibinfo{pages}{709} (\bibinfo{year}{2012}{\natexlab{a}}).

\bibitem[{\citenamefont{Fernandes et~al.}(2014)\citenamefont{Fernandes,
  Chubukov, and Schmalian}}]{Fernandes2014}
\bibinfo{author}{\bibfnamefont{R.~M.} \bibnamefont{Fernandes}},
  \bibinfo{author}{\bibfnamefont{A.~V.} \bibnamefont{Chubukov}},
  \bibnamefont{and}
  \bibinfo{author}{\bibfnamefont{J.}~\bibnamefont{Schmalian}},
  \bibinfo{journal}{Nat. Phys.} \textbf{\bibinfo{volume}{10}},
  \bibinfo{pages}{97} (\bibinfo{year}{2014}).

\bibitem[{\citenamefont{Chu et~al.}(2010)\citenamefont{Chu, Analytis, {De
  Greve}, McMahon, Islam, Yamamoto, and Fisher}}]{Chu2010}
\bibinfo{author}{\bibfnamefont{J.-H.} \bibnamefont{Chu}},
  \bibinfo{author}{\bibfnamefont{J.~G.} \bibnamefont{Analytis}},
  \bibinfo{author}{\bibfnamefont{K.}~\bibnamefont{{De Greve}}},
  \bibinfo{author}{\bibfnamefont{P.~L.} \bibnamefont{McMahon}},
  \bibinfo{author}{\bibfnamefont{Z.}~\bibnamefont{Islam}},
  \bibinfo{author}{\bibfnamefont{Y.}~\bibnamefont{Yamamoto}}, \bibnamefont{and}
  \bibinfo{author}{\bibfnamefont{I.~R.} \bibnamefont{Fisher}},
  \bibinfo{journal}{Science} \textbf{\bibinfo{volume}{329}},
  \bibinfo{pages}{824} (\bibinfo{year}{2010}).

\bibitem[{\citenamefont{Yi et~al.}(2011)\citenamefont{Yi, Lu, Chu, Analytis,
  Sorini, Kemper, Moritz, Mo, Moore, Hashimoto et~al.}}]{Yi2011}
\bibinfo{author}{\bibfnamefont{M.}~\bibnamefont{Yi}},
  \bibinfo{author}{\bibfnamefont{D.}~\bibnamefont{Lu}},
  \bibinfo{author}{\bibfnamefont{J.~J.-H.} \bibnamefont{Chu}},
  \bibinfo{author}{\bibfnamefont{J.~G.} \bibnamefont{Analytis}},
  \bibinfo{author}{\bibfnamefont{A.~P.} \bibnamefont{Sorini}},
  \bibinfo{author}{\bibfnamefont{A.~F.} \bibnamefont{Kemper}},
  \bibinfo{author}{\bibfnamefont{B.}~\bibnamefont{Moritz}},
  \bibinfo{author}{\bibfnamefont{S.-K.} \bibnamefont{Mo}},
  \bibinfo{author}{\bibfnamefont{R.~G.} \bibnamefont{Moore}},
  \bibinfo{author}{\bibfnamefont{M.}~\bibnamefont{Hashimoto}},
  \bibnamefont{et~al.}, \bibinfo{journal}{Proc. Natl. Acad. Sci.}
  \textbf{\bibinfo{volume}{108}}, \bibinfo{pages}{6878} (\bibinfo{year}{2011}).

\bibitem[{\citenamefont{Tanatar et~al.}(2010)\citenamefont{Tanatar, Blomberg,
  Kreyssig, Kim, Ni, Thaler, Bud'ko, Canfield, Goldman, Mazin
  et~al.}}]{Tanatar2010}
\bibinfo{author}{\bibfnamefont{M.~A.} \bibnamefont{Tanatar}},
  \bibinfo{author}{\bibfnamefont{E.~C.} \bibnamefont{Blomberg}},
  \bibinfo{author}{\bibfnamefont{A.}~\bibnamefont{Kreyssig}},
  \bibinfo{author}{\bibfnamefont{M.~G.} \bibnamefont{Kim}},
  \bibinfo{author}{\bibfnamefont{N.}~\bibnamefont{Ni}},
  \bibinfo{author}{\bibfnamefont{A.}~\bibnamefont{Thaler}},
  \bibinfo{author}{\bibfnamefont{S.~L.} \bibnamefont{Bud'ko}},
  \bibinfo{author}{\bibfnamefont{P.~C.} \bibnamefont{Canfield}},
  \bibinfo{author}{\bibfnamefont{A.~I.} \bibnamefont{Goldman}},
  \bibinfo{author}{\bibfnamefont{I.~I.} \bibnamefont{Mazin}},
  \bibnamefont{et~al.}, \bibinfo{journal}{Phys. Rev. B}
  \textbf{\bibinfo{volume}{81}}, \bibinfo{pages}{184508}
  (\bibinfo{year}{2010}).

\bibitem[{\citenamefont{Kasahara et~al.}(2012)\citenamefont{Kasahara, Shi,
  Hashimoto, Tonegawa, Mizukami, Shibauchi, Sugimoto, Fukuda, Terashima,
  Nevidomskyy et~al.}}]{Kasahara2012}
\bibinfo{author}{\bibfnamefont{S.}~\bibnamefont{Kasahara}},
  \bibinfo{author}{\bibfnamefont{H.~J.} \bibnamefont{Shi}},
  \bibinfo{author}{\bibfnamefont{K.}~\bibnamefont{Hashimoto}},
  \bibinfo{author}{\bibfnamefont{S.}~\bibnamefont{Tonegawa}},
  \bibinfo{author}{\bibfnamefont{Y.}~\bibnamefont{Mizukami}},
  \bibinfo{author}{\bibfnamefont{T.}~\bibnamefont{Shibauchi}},
  \bibinfo{author}{\bibfnamefont{K.}~\bibnamefont{Sugimoto}},
  \bibinfo{author}{\bibfnamefont{T.}~\bibnamefont{Fukuda}},
  \bibinfo{author}{\bibfnamefont{T.}~\bibnamefont{Terashima}},
  \bibinfo{author}{\bibfnamefont{A.~H.} \bibnamefont{Nevidomskyy}},
  \bibnamefont{et~al.}, \bibinfo{journal}{Nature}
  \textbf{\bibinfo{volume}{486}}, \bibinfo{pages}{382} (\bibinfo{year}{2012}).

\bibitem[{\citenamefont{Fernandes and Schmalian}(2012)}]{Fernandes2012}
\bibinfo{author}{\bibfnamefont{R.~M.} \bibnamefont{Fernandes}}
  \bibnamefont{and}
  \bibinfo{author}{\bibfnamefont{J.}~\bibnamefont{Schmalian}},
  \bibinfo{journal}{Supercond. Sci. Technol.} \textbf{\bibinfo{volume}{25}},
  \bibinfo{pages}{84005} (\bibinfo{year}{2012}).

\bibitem[{\citenamefont{Nakajima et~al.}(2011)\citenamefont{Nakajima, Liang,
  Ishida, Tomioka, Kihou, Lee, Iyo, Eisaki, Kakeshita, Ito
  et~al.}}]{Nakajima2011a}
\bibinfo{author}{\bibfnamefont{M.}~\bibnamefont{Nakajima}},
  \bibinfo{author}{\bibfnamefont{T.}~\bibnamefont{Liang}},
  \bibinfo{author}{\bibfnamefont{S.}~\bibnamefont{Ishida}},
  \bibinfo{author}{\bibfnamefont{Y.}~\bibnamefont{Tomioka}},
  \bibinfo{author}{\bibfnamefont{K.}~\bibnamefont{Kihou}},
  \bibinfo{author}{\bibfnamefont{C.~H.} \bibnamefont{Lee}},
  \bibinfo{author}{\bibfnamefont{A.}~\bibnamefont{Iyo}},
  \bibinfo{author}{\bibfnamefont{H.}~\bibnamefont{Eisaki}},
  \bibinfo{author}{\bibfnamefont{T.}~\bibnamefont{Kakeshita}},
  \bibinfo{author}{\bibfnamefont{T.}~\bibnamefont{Ito}}, \bibnamefont{et~al.},
  \bibinfo{journal}{Proc. Natl. Acad. Sci. U. S. A.}
  \textbf{\bibinfo{volume}{108}}, \bibinfo{pages}{12238}
  (\bibinfo{year}{2011}).

\bibitem[{\citenamefont{Nakajima et~al.}(2012)\citenamefont{Nakajima, Ishida,
  Tomioka, Kihou, Lee, Iyo, Ito, Kakeshita, Eisaki, and Uchida}}]{Nakajima2012}
\bibinfo{author}{\bibfnamefont{M.}~\bibnamefont{Nakajima}},
  \bibinfo{author}{\bibfnamefont{S.}~\bibnamefont{Ishida}},
  \bibinfo{author}{\bibfnamefont{Y.}~\bibnamefont{Tomioka}},
  \bibinfo{author}{\bibfnamefont{K.}~\bibnamefont{Kihou}},
  \bibinfo{author}{\bibfnamefont{C.~H.} \bibnamefont{Lee}},
  \bibinfo{author}{\bibfnamefont{A.}~\bibnamefont{Iyo}},
  \bibinfo{author}{\bibfnamefont{T.}~\bibnamefont{Ito}},
  \bibinfo{author}{\bibfnamefont{T.}~\bibnamefont{Kakeshita}},
  \bibinfo{author}{\bibfnamefont{H.}~\bibnamefont{Eisaki}}, \bibnamefont{and}
  \bibinfo{author}{\bibfnamefont{S.}~\bibnamefont{Uchida}},
  \bibinfo{journal}{Phys. Rev. Lett.} \textbf{\bibinfo{volume}{109}},
  \bibinfo{pages}{217003} (\bibinfo{year}{2012}).

\bibitem[{\citenamefont{Dusza et~al.}(2011)\citenamefont{Dusza, Lucarelli,
  Pfuner, Chu, Fisher, and Degiorgi}}]{Dusza2011a}
\bibinfo{author}{\bibfnamefont{A.}~\bibnamefont{Dusza}},
  \bibinfo{author}{\bibfnamefont{A.}~\bibnamefont{Lucarelli}},
  \bibinfo{author}{\bibfnamefont{F.}~\bibnamefont{Pfuner}},
  \bibinfo{author}{\bibfnamefont{J.-H.} \bibnamefont{Chu}},
  \bibinfo{author}{\bibfnamefont{I.~R.} \bibnamefont{Fisher}},
  \bibnamefont{and} \bibinfo{author}{\bibfnamefont{L.}~\bibnamefont{Degiorgi}},
  \bibinfo{journal}{EPL (Europhysics Lett.} \textbf{\bibinfo{volume}{93}},
  \bibinfo{pages}{37002} (\bibinfo{year}{2011}).

\bibitem[{\citenamefont{Kr\"uger et~al.}(2009)\citenamefont{Kr\"uger, Kumar,
  Zaanen, and van~den Brink}}]{Kruger2009}
\bibinfo{author}{\bibfnamefont{F.}~\bibnamefont{Kr\"uger}},
  \bibinfo{author}{\bibfnamefont{S.}~\bibnamefont{Kumar}},
  \bibinfo{author}{\bibfnamefont{J.}~\bibnamefont{Zaanen}}, \bibnamefont{and}
  \bibinfo{author}{\bibfnamefont{J.}~\bibnamefont{van~den Brink}},
  \bibinfo{journal}{Phys. Rev. B} \textbf{\bibinfo{volume}{79}},
  \bibinfo{pages}{054504} (\bibinfo{year}{2009}).

\bibitem[{\citenamefont{Lv et~al.}(2009)\citenamefont{Lv, Wu, and
  Phillips}}]{Lv2009}
\bibinfo{author}{\bibfnamefont{W.}~\bibnamefont{Lv}},
  \bibinfo{author}{\bibfnamefont{J.}~\bibnamefont{Wu}}, \bibnamefont{and}
  \bibinfo{author}{\bibfnamefont{P.}~\bibnamefont{Phillips}},
  \bibinfo{journal}{Phys. Rev. B} \textbf{\bibinfo{volume}{80}},
  \bibinfo{pages}{224506} (\bibinfo{year}{2009}).

\bibitem[{\citenamefont{Chen et~al.}(2010)\citenamefont{Chen, Maciejko, Sorini,
  Moritz, Singh, and Devereaux}}]{Chen2010}
\bibinfo{author}{\bibfnamefont{C.-C.} \bibnamefont{Chen}},
  \bibinfo{author}{\bibfnamefont{J.}~\bibnamefont{Maciejko}},
  \bibinfo{author}{\bibfnamefont{A.~P.} \bibnamefont{Sorini}},
  \bibinfo{author}{\bibfnamefont{B.}~\bibnamefont{Moritz}},
  \bibinfo{author}{\bibfnamefont{R.~R.~P.} \bibnamefont{Singh}},
  \bibnamefont{and} \bibinfo{author}{\bibfnamefont{T.~P.}
  \bibnamefont{Devereaux}}, \bibinfo{journal}{Phys. Rev. B}
  \textbf{\bibinfo{volume}{82}}, \bibinfo{pages}{100504}
  (\bibinfo{year}{2010}).

\bibitem[{\citenamefont{Eremin and Chubukov}(2010)}]{Eremin2010}
\bibinfo{author}{\bibfnamefont{I.}~\bibnamefont{Eremin}} \bibnamefont{and}
  \bibinfo{author}{\bibfnamefont{A.~V.} \bibnamefont{Chubukov}},
  \bibinfo{journal}{Phys. Rev. B} \textbf{\bibinfo{volume}{81}},
  \bibinfo{pages}{024511} (\bibinfo{year}{2010}).

\bibitem[{\citenamefont{Xu et~al.}(2008)\citenamefont{Xu, M\"uller, and
  Sachdev}}]{Xu2008}
\bibinfo{author}{\bibfnamefont{C.}~\bibnamefont{Xu}},
  \bibinfo{author}{\bibfnamefont{M.}~\bibnamefont{M\"uller}}, \bibnamefont{and}
  \bibinfo{author}{\bibfnamefont{S.}~\bibnamefont{Sachdev}},
  \bibinfo{journal}{Phys. Rev. B} \textbf{\bibinfo{volume}{78}},
  \bibinfo{pages}{020501} (\bibinfo{year}{2008}).

\bibitem[{\citenamefont{Fernandes et~al.}(2012)\citenamefont{Fernandes,
  Chubukov, Knolle, Eremin, and Schmalian}}]{Fernandes2012PRB}
\bibinfo{author}{\bibfnamefont{R.~M.} \bibnamefont{Fernandes}},
  \bibinfo{author}{\bibfnamefont{A.~V.} \bibnamefont{Chubukov}},
  \bibinfo{author}{\bibfnamefont{J.}~\bibnamefont{Knolle}},
  \bibinfo{author}{\bibfnamefont{I.}~\bibnamefont{Eremin}}, \bibnamefont{and}
  \bibinfo{author}{\bibfnamefont{J.}~\bibnamefont{Schmalian}},
  \bibinfo{journal}{Phys. Rev. B} \textbf{\bibinfo{volume}{85}},
  \bibinfo{pages}{024534} (\bibinfo{year}{2012}).

\bibitem[{\citenamefont{Avci et~al.}(2014)\citenamefont{Avci, Chmaissem,
  Allred, Rosenkranz, Eremin, Chubukov, Bugaris, Chung, Kanatzidis, Castellan
  et~al.}}]{Avci2014}
\bibinfo{author}{\bibfnamefont{S.}~\bibnamefont{Avci}},
  \bibinfo{author}{\bibfnamefont{O.}~\bibnamefont{Chmaissem}},
  \bibinfo{author}{\bibfnamefont{J.~M.} \bibnamefont{Allred}},
  \bibinfo{author}{\bibfnamefont{S.}~\bibnamefont{Rosenkranz}},
  \bibinfo{author}{\bibfnamefont{I.}~\bibnamefont{Eremin}},
  \bibinfo{author}{\bibfnamefont{a.~V.} \bibnamefont{Chubukov}},
  \bibinfo{author}{\bibfnamefont{D.~E.} \bibnamefont{Bugaris}},
  \bibinfo{author}{\bibfnamefont{D.~Y.} \bibnamefont{Chung}},
  \bibinfo{author}{\bibfnamefont{M.~G.} \bibnamefont{Kanatzidis}},
  \bibinfo{author}{\bibfnamefont{J.-P.} \bibnamefont{Castellan}},
  \bibnamefont{et~al.}, \bibinfo{journal}{Nat. Commun.}
  \textbf{\bibinfo{volume}{5}}, \bibinfo{pages}{3845} (\bibinfo{year}{2014}).

\bibitem[{\citenamefont{Mazin et~al.}(2008)\citenamefont{Mazin, Singh,
  Johannes, and Du}}]{Mazin2008}
\bibinfo{author}{\bibfnamefont{I.~I.} \bibnamefont{Mazin}},
  \bibinfo{author}{\bibfnamefont{D.~J.} \bibnamefont{Singh}},
  \bibinfo{author}{\bibfnamefont{M.~D.} \bibnamefont{Johannes}},
  \bibnamefont{and} \bibinfo{author}{\bibfnamefont{M.~H.} \bibnamefont{Du}},
  \bibinfo{journal}{Phys. Rev. Lett.} \textbf{\bibinfo{volume}{101}},
  \bibinfo{pages}{057003} (\bibinfo{year}{2008}).

\bibitem[{\citenamefont{Kuroki et~al.}(2008)\citenamefont{Kuroki, Onari, Arita,
  Usui, Tanaka, Kontani, and Aoki}}]{Kuroki2008}
\bibinfo{author}{\bibfnamefont{K.}~\bibnamefont{Kuroki}},
  \bibinfo{author}{\bibfnamefont{S.}~\bibnamefont{Onari}},
  \bibinfo{author}{\bibfnamefont{R.}~\bibnamefont{Arita}},
  \bibinfo{author}{\bibfnamefont{H.}~\bibnamefont{Usui}},
  \bibinfo{author}{\bibfnamefont{Y.}~\bibnamefont{Tanaka}},
  \bibinfo{author}{\bibfnamefont{H.}~\bibnamefont{Kontani}}, \bibnamefont{and}
  \bibinfo{author}{\bibfnamefont{H.}~\bibnamefont{Aoki}},
  \bibinfo{journal}{Phys. Rev. Lett.} \textbf{\bibinfo{volume}{101}},
  \bibinfo{pages}{087004} (\bibinfo{year}{2008}).

\bibitem[{\citenamefont{Kontani and Onari}(2010)}]{Kontani2010}
\bibinfo{author}{\bibfnamefont{H.}~\bibnamefont{Kontani}} \bibnamefont{and}
  \bibinfo{author}{\bibfnamefont{S.}~\bibnamefont{Onari}},
  \bibinfo{journal}{Phys. Rev. Lett.} \textbf{\bibinfo{volume}{104}},
  \bibinfo{pages}{157001} (\bibinfo{year}{2010}).

\bibitem[{\citenamefont{Kontani et~al.}(2011)\citenamefont{Kontani, Saito, and
  Onari}}]{Kontani2011}
\bibinfo{author}{\bibfnamefont{H.}~\bibnamefont{Kontani}},
  \bibinfo{author}{\bibfnamefont{T.}~\bibnamefont{Saito}}, \bibnamefont{and}
  \bibinfo{author}{\bibfnamefont{S.}~\bibnamefont{Onari}},
  \bibinfo{journal}{Phys. Rev. B} \textbf{\bibinfo{volume}{84}},
  \bibinfo{pages}{024528} (\bibinfo{year}{2011}).

\bibitem[{\citenamefont{Hu et~al.}(2008)\citenamefont{Hu, Dong, Li, Li, Zheng,
  Chen, Luo, and Wang}}]{Hu2008}
\bibinfo{author}{\bibfnamefont{W.~Z.} \bibnamefont{Hu}},
  \bibinfo{author}{\bibfnamefont{J.}~\bibnamefont{Dong}},
  \bibinfo{author}{\bibfnamefont{G.}~\bibnamefont{Li}},
  \bibinfo{author}{\bibfnamefont{Z.}~\bibnamefont{Li}},
  \bibinfo{author}{\bibfnamefont{P.}~\bibnamefont{Zheng}},
  \bibinfo{author}{\bibfnamefont{G.~F.} \bibnamefont{Chen}},
  \bibinfo{author}{\bibfnamefont{J.~L.} \bibnamefont{Luo}}, \bibnamefont{and}
  \bibinfo{author}{\bibfnamefont{N.~L.} \bibnamefont{Wang}},
  \bibinfo{journal}{Phys. Rev. Lett.} \textbf{\bibinfo{volume}{101}},
  \bibinfo{pages}{257005} (\bibinfo{year}{2008}).

\bibitem[{\citenamefont{Li et~al.}(2008)\citenamefont{Li, Hu, Dong, Li, Zheng,
  Chen, Luo, and Wang}}]{Li2008}
\bibinfo{author}{\bibfnamefont{G.}~\bibnamefont{Li}},
  \bibinfo{author}{\bibfnamefont{W.~Z.} \bibnamefont{Hu}},
  \bibinfo{author}{\bibfnamefont{J.}~\bibnamefont{Dong}},
  \bibinfo{author}{\bibfnamefont{Z.}~\bibnamefont{Li}},
  \bibinfo{author}{\bibfnamefont{P.}~\bibnamefont{Zheng}},
  \bibinfo{author}{\bibfnamefont{G.~F.} \bibnamefont{Chen}},
  \bibinfo{author}{\bibfnamefont{J.~L.} \bibnamefont{Luo}}, \bibnamefont{and}
  \bibinfo{author}{\bibfnamefont{N.~L.} \bibnamefont{Wang}},
  \bibinfo{journal}{Phys. Rev. Lett.} \textbf{\bibinfo{volume}{101}},
  \bibinfo{pages}{107004} (\bibinfo{year}{2008}).

\bibitem[{\citenamefont{Lobo et~al.}(2010)\citenamefont{Lobo, Dai, Nagel, R\~o\
  om, Carbotte, Timusk, Forget, and Colson}}]{Lobo2010}
\bibinfo{author}{\bibfnamefont{R.~P. S.~M.} \bibnamefont{Lobo}},
  \bibinfo{author}{\bibfnamefont{Y.~M.} \bibnamefont{Dai}},
  \bibinfo{author}{\bibfnamefont{U.}~\bibnamefont{Nagel}},
  \bibinfo{author}{\bibfnamefont{T.}~\bibnamefont{R\~o\ om}},
  \bibinfo{author}{\bibfnamefont{J.~P.} \bibnamefont{Carbotte}},
  \bibinfo{author}{\bibfnamefont{T.}~\bibnamefont{Timusk}},
  \bibinfo{author}{\bibfnamefont{A.}~\bibnamefont{Forget}}, \bibnamefont{and}
  \bibinfo{author}{\bibfnamefont{D.}~\bibnamefont{Colson}},
  \bibinfo{journal}{Phys. Rev. B} \textbf{\bibinfo{volume}{82}},
  \bibinfo{pages}{100506} (\bibinfo{year}{2010}).

\bibitem[{\citenamefont{Dai et~al.}(2012{\natexlab{b}})\citenamefont{Dai, Xu,
  Shen, Wen, Hu, Qiu, and Lobo}}]{Dai2012PRB}
\bibinfo{author}{\bibfnamefont{Y.~M.} \bibnamefont{Dai}},
  \bibinfo{author}{\bibfnamefont{B.}~\bibnamefont{Xu}},
  \bibinfo{author}{\bibfnamefont{B.}~\bibnamefont{Shen}},
  \bibinfo{author}{\bibfnamefont{H.~H.} \bibnamefont{Wen}},
  \bibinfo{author}{\bibfnamefont{J.~P.} \bibnamefont{Hu}},
  \bibinfo{author}{\bibfnamefont{X.~G.} \bibnamefont{Qiu}}, \bibnamefont{and}
  \bibinfo{author}{\bibfnamefont{R.~P. S.~M.} \bibnamefont{Lobo}},
  \bibinfo{journal}{Phys. Rev. B} \textbf{\bibinfo{volume}{86}},
  \bibinfo{pages}{100501} (\bibinfo{year}{2012}{\natexlab{b}}).

\bibitem[{\citenamefont{Dai et~al.}(2013)\citenamefont{Dai, Xu, Shen, Wen, Qiu,
  Lobo, and {S. M. Lobo}}}]{Dai2013a}
\bibinfo{author}{\bibfnamefont{Y.~M.} \bibnamefont{Dai}},
  \bibinfo{author}{\bibfnamefont{B.}~\bibnamefont{Xu}},
  \bibinfo{author}{\bibfnamefont{B.}~\bibnamefont{Shen}},
  \bibinfo{author}{\bibfnamefont{H.-H.} \bibnamefont{Wen}},
  \bibinfo{author}{\bibfnamefont{X.~G.} \bibnamefont{Qiu}},
  \bibinfo{author}{\bibfnamefont{R.~P. S.~M.} \bibnamefont{Lobo}},
  \bibnamefont{and} \bibinfo{author}{\bibfnamefont{R.~P.} \bibnamefont{{S. M.
  Lobo}}}, \bibinfo{journal}{EPL (Europhysics Letters)}
  \textbf{\bibinfo{volume}{104}}, \bibinfo{pages}{47006}
  (\bibinfo{year}{2013}).

\bibitem[{\citenamefont{Dai et~al.}(2016)\citenamefont{Dai, Miao, Xing, Wang,
  Jin, Ding, and Homes}}]{Dai2016}
\bibinfo{author}{\bibfnamefont{Y.~M.} \bibnamefont{Dai}},
  \bibinfo{author}{\bibfnamefont{H.}~\bibnamefont{Miao}},
  \bibinfo{author}{\bibfnamefont{L.~Y.} \bibnamefont{Xing}},
  \bibinfo{author}{\bibfnamefont{X.~C.} \bibnamefont{Wang}},
  \bibinfo{author}{\bibfnamefont{C.~Q.} \bibnamefont{Jin}},
  \bibinfo{author}{\bibfnamefont{H.}~\bibnamefont{Ding}}, \bibnamefont{and}
  \bibinfo{author}{\bibfnamefont{C.~C.} \bibnamefont{Homes}},
  \bibinfo{journal}{Phys. Rev. B} \textbf{\bibinfo{volume}{93}},
  \bibinfo{pages}{054508} (\bibinfo{year}{2016}).

\bibitem[{\citenamefont{Blomberg et~al.}(2013)\citenamefont{Blomberg, Tanatar,
  Fernandes, Mazin, Shen, Wen, Johannes, Schmalian, and
  Prozorov}}]{Blomberg2013}
\bibinfo{author}{\bibfnamefont{E.~C.} \bibnamefont{Blomberg}},
  \bibinfo{author}{\bibfnamefont{M.~a.} \bibnamefont{Tanatar}},
  \bibinfo{author}{\bibfnamefont{R.~M.} \bibnamefont{Fernandes}},
  \bibinfo{author}{\bibfnamefont{I.~I.} \bibnamefont{Mazin}},
  \bibinfo{author}{\bibfnamefont{B.}~\bibnamefont{Shen}},
  \bibinfo{author}{\bibfnamefont{H.-H.} \bibnamefont{Wen}},
  \bibinfo{author}{\bibfnamefont{M.~D.} \bibnamefont{Johannes}},
  \bibinfo{author}{\bibfnamefont{J.}~\bibnamefont{Schmalian}},
  \bibnamefont{and} \bibinfo{author}{\bibfnamefont{R.}~\bibnamefont{Prozorov}},
  \bibinfo{journal}{Nat. Commun.} \textbf{\bibinfo{volume}{4}},
  \bibinfo{pages}{1914} (\bibinfo{year}{2013}).

\bibitem[{\citenamefont{Shen et~al.}(2011)\citenamefont{Shen, Yang, Wang, Han,
  Zeng, Shan, Ren, and Wen}}]{Shen2011}
\bibinfo{author}{\bibfnamefont{B.}~\bibnamefont{Shen}},
  \bibinfo{author}{\bibfnamefont{H.}~\bibnamefont{Yang}},
  \bibinfo{author}{\bibfnamefont{Z.-S.} \bibnamefont{Wang}},
  \bibinfo{author}{\bibfnamefont{F.}~\bibnamefont{Han}},
  \bibinfo{author}{\bibfnamefont{B.}~\bibnamefont{Zeng}},
  \bibinfo{author}{\bibfnamefont{L.}~\bibnamefont{Shan}},
  \bibinfo{author}{\bibfnamefont{C.}~\bibnamefont{Ren}}, \bibnamefont{and}
  \bibinfo{author}{\bibfnamefont{H.-H.} \bibnamefont{Wen}},
  \bibinfo{journal}{Phys. Rev. B} \textbf{\bibinfo{volume}{84}},
  \bibinfo{pages}{184512} (\bibinfo{year}{2011}).

\bibitem[{\citenamefont{Rullier-Albenque
  et~al.}(2009)\citenamefont{Rullier-Albenque, Colson, Forget, and
  Alloul}}]{Rullier-Albenque2009}
\bibinfo{author}{\bibfnamefont{F.}~\bibnamefont{Rullier-Albenque}},
  \bibinfo{author}{\bibfnamefont{D.}~\bibnamefont{Colson}},
  \bibinfo{author}{\bibfnamefont{A.}~\bibnamefont{Forget}}, \bibnamefont{and}
  \bibinfo{author}{\bibfnamefont{H.}~\bibnamefont{Alloul}},
  \bibinfo{journal}{Phys. Rev. Lett.} \textbf{\bibinfo{volume}{103}},
  \bibinfo{pages}{057001} (\bibinfo{year}{2009}).

\bibitem[{\citenamefont{Ye et~al.}(2012)\citenamefont{Ye, Zhang, Chen, Xu, Ge,
  Jiang, Xie, and Feng}}]{Ye2012}
\bibinfo{author}{\bibfnamefont{Z.~R.} \bibnamefont{Ye}},
  \bibinfo{author}{\bibfnamefont{Y.}~\bibnamefont{Zhang}},
  \bibinfo{author}{\bibfnamefont{F.}~\bibnamefont{Chen}},
  \bibinfo{author}{\bibfnamefont{M.}~\bibnamefont{Xu}},
  \bibinfo{author}{\bibfnamefont{Q.~Q.} \bibnamefont{Ge}},
  \bibinfo{author}{\bibfnamefont{J.}~\bibnamefont{Jiang}},
  \bibinfo{author}{\bibfnamefont{B.~P.} \bibnamefont{Xie}}, \bibnamefont{and}
  \bibinfo{author}{\bibfnamefont{D.~L.} \bibnamefont{Feng}},
  \bibinfo{journal}{Phys. Rev. B} \textbf{\bibinfo{volume}{86}},
  \bibinfo{pages}{035136} (\bibinfo{year}{2012}).

\bibitem[{\citenamefont{Homes et~al.}(1993)\citenamefont{Homes, Reedyk,
  Cradles, and Timusk}}]{Homes1993}
\bibinfo{author}{\bibfnamefont{C.~C.} \bibnamefont{Homes}},
  \bibinfo{author}{\bibfnamefont{M.}~\bibnamefont{Reedyk}},
  \bibinfo{author}{\bibfnamefont{D.~A.} \bibnamefont{Cradles}},
  \bibnamefont{and} \bibinfo{author}{\bibfnamefont{T.}~\bibnamefont{Timusk}},
  \bibinfo{journal}{Appl. Opt.} \textbf{\bibinfo{volume}{32}},
  \bibinfo{pages}{2976} (\bibinfo{year}{1993}).

\bibitem[{\citenamefont{Wilson et~al.}(2010)\citenamefont{Wilson, Yamani,
  Rotundu, Freelon, Valdivia, Bourret-Courchesne, Lynn, Chi, Hong, and
  Birgeneau}}]{Wilson2010}
\bibinfo{author}{\bibfnamefont{S.~D.} \bibnamefont{Wilson}},
  \bibinfo{author}{\bibfnamefont{Z.}~\bibnamefont{Yamani}},
  \bibinfo{author}{\bibfnamefont{C.~R.} \bibnamefont{Rotundu}},
  \bibinfo{author}{\bibfnamefont{B.}~\bibnamefont{Freelon}},
  \bibinfo{author}{\bibfnamefont{P.~N.} \bibnamefont{Valdivia}},
  \bibinfo{author}{\bibfnamefont{E.}~\bibnamefont{Bourret-Courchesne}},
  \bibinfo{author}{\bibfnamefont{J.~W.} \bibnamefont{Lynn}},
  \bibinfo{author}{\bibfnamefont{S.}~\bibnamefont{Chi}},
  \bibinfo{author}{\bibfnamefont{T.}~\bibnamefont{Hong}}, \bibnamefont{and}
  \bibinfo{author}{\bibfnamefont{R.~J.} \bibnamefont{Birgeneau}},
  \bibinfo{journal}{Phys. Rev. B} \textbf{\bibinfo{volume}{82}},
  \bibinfo{pages}{144502} (\bibinfo{year}{2010}).

\bibitem[{\citenamefont{Ning et~al.}(2010)\citenamefont{Ning, Ahilan, Imai,
  Sefat, McGuire, Sales, Mandrus, Cheng, Shen, and Wen}}]{Ning2010}
\bibinfo{author}{\bibfnamefont{F.~L.} \bibnamefont{Ning}},
  \bibinfo{author}{\bibfnamefont{K.}~\bibnamefont{Ahilan}},
  \bibinfo{author}{\bibfnamefont{T.}~\bibnamefont{Imai}},
  \bibinfo{author}{\bibfnamefont{A.~S.} \bibnamefont{Sefat}},
  \bibinfo{author}{\bibfnamefont{M.~A.} \bibnamefont{McGuire}},
  \bibinfo{author}{\bibfnamefont{B.~C.} \bibnamefont{Sales}},
  \bibinfo{author}{\bibfnamefont{D.}~\bibnamefont{Mandrus}},
  \bibinfo{author}{\bibfnamefont{P.}~\bibnamefont{Cheng}},
  \bibinfo{author}{\bibfnamefont{B.}~\bibnamefont{Shen}}, \bibnamefont{and}
  \bibinfo{author}{\bibfnamefont{H.-H.} \bibnamefont{Wen}},
  \bibinfo{journal}{Phys. Rev. Lett.} \textbf{\bibinfo{volume}{104}},
  \bibinfo{pages}{037001} (\bibinfo{year}{2010}).

\bibitem[{\citenamefont{Nakai et~al.}(2010)\citenamefont{Nakai, Iye, Kitagawa,
  Ishida, Ikeda, Kasahara, Shishido, Shibauchi, Matsuda, and
  Terashima}}]{Nakai2010b}
\bibinfo{author}{\bibfnamefont{Y.}~\bibnamefont{Nakai}},
  \bibinfo{author}{\bibfnamefont{T.}~\bibnamefont{Iye}},
  \bibinfo{author}{\bibfnamefont{S.}~\bibnamefont{Kitagawa}},
  \bibinfo{author}{\bibfnamefont{K.}~\bibnamefont{Ishida}},
  \bibinfo{author}{\bibfnamefont{H.}~\bibnamefont{Ikeda}},
  \bibinfo{author}{\bibfnamefont{S.}~\bibnamefont{Kasahara}},
  \bibinfo{author}{\bibfnamefont{H.}~\bibnamefont{Shishido}},
  \bibinfo{author}{\bibfnamefont{T.}~\bibnamefont{Shibauchi}},
  \bibinfo{author}{\bibfnamefont{Y.}~\bibnamefont{Matsuda}}, \bibnamefont{and}
  \bibinfo{author}{\bibfnamefont{T.}~\bibnamefont{Terashima}},
  \bibinfo{journal}{Phys. Rev. Lett.} \textbf{\bibinfo{volume}{105}},
  \bibinfo{pages}{107003} (\bibinfo{year}{2010}).

\bibitem[{\citenamefont{Li et~al.}(2011)\citenamefont{Li, Sun, Lin, Su, Hu, and
  Zheng}}]{Li2011}
\bibinfo{author}{\bibfnamefont{Z.}~\bibnamefont{Li}},
  \bibinfo{author}{\bibfnamefont{D.~L.} \bibnamefont{Sun}},
  \bibinfo{author}{\bibfnamefont{C.~T.} \bibnamefont{Lin}},
  \bibinfo{author}{\bibfnamefont{Y.~H.} \bibnamefont{Su}},
  \bibinfo{author}{\bibfnamefont{J.~P.} \bibnamefont{Hu}}, \bibnamefont{and}
  \bibinfo{author}{\bibfnamefont{G.-q.} \bibnamefont{Zheng}},
  \bibinfo{journal}{Phys. Rev. B} \textbf{\bibinfo{volume}{83}},
  \bibinfo{pages}{140506} (\bibinfo{year}{2011}).

\bibitem[{\citenamefont{Moon et~al.}(2014)\citenamefont{Moon, Lee, Schafgans,
  Chubukov, Kasahara, Shibauchi, Terashima, Matsuda, Tanatar, Prozorov
  et~al.}}]{Moon2014}
\bibinfo{author}{\bibfnamefont{S.~J.} \bibnamefont{Moon}},
  \bibinfo{author}{\bibfnamefont{Y.~S.} \bibnamefont{Lee}},
  \bibinfo{author}{\bibfnamefont{A.~A.} \bibnamefont{Schafgans}},
  \bibinfo{author}{\bibfnamefont{A.~V.} \bibnamefont{Chubukov}},
  \bibinfo{author}{\bibfnamefont{S.}~\bibnamefont{Kasahara}},
  \bibinfo{author}{\bibfnamefont{T.}~\bibnamefont{Shibauchi}},
  \bibinfo{author}{\bibfnamefont{T.}~\bibnamefont{Terashima}},
  \bibinfo{author}{\bibfnamefont{Y.}~\bibnamefont{Matsuda}},
  \bibinfo{author}{\bibfnamefont{M.~A.} \bibnamefont{Tanatar}},
  \bibinfo{author}{\bibfnamefont{R.}~\bibnamefont{Prozorov}},
  \bibnamefont{et~al.}, \bibinfo{journal}{Phys. Rev. B}
  \textbf{\bibinfo{volume}{90}}, \bibinfo{pages}{014503}
  (\bibinfo{year}{2014}).

\bibitem[{\citenamefont{Yin et~al.}(2011)\citenamefont{Yin, Haule, and
  Kotliar}}]{Yin2011NP}
\bibinfo{author}{\bibfnamefont{Z.~P.} \bibnamefont{Yin}},
  \bibinfo{author}{\bibfnamefont{K.}~\bibnamefont{Haule}}, \bibnamefont{and}
  \bibinfo{author}{\bibfnamefont{G.}~\bibnamefont{Kotliar}},
  \bibinfo{journal}{Nat. Phys.} \textbf{\bibinfo{volume}{7}},
  \bibinfo{pages}{294} (\bibinfo{year}{2011}).

\bibitem[{\citenamefont{B{\"{o}}hmer et~al.}(2015)\citenamefont{B{\"{o}}hmer,
  Hardy, Wang, Wolf, Schweiss, and Meingast}}]{Boehmer2015}
\bibinfo{author}{\bibfnamefont{A.~E.} \bibnamefont{B{\"{o}}hmer}},
  \bibinfo{author}{\bibfnamefont{F.}~\bibnamefont{Hardy}},
  \bibinfo{author}{\bibfnamefont{L.}~\bibnamefont{Wang}},
  \bibinfo{author}{\bibfnamefont{T.}~\bibnamefont{Wolf}},
  \bibinfo{author}{\bibfnamefont{P.}~\bibnamefont{Schweiss}}, \bibnamefont{and}
  \bibinfo{author}{\bibfnamefont{C.}~\bibnamefont{Meingast}},
  \bibinfo{journal}{Nat. Commun.} \textbf{\bibinfo{volume}{6}},
  \bibinfo{pages}{7911} (\bibinfo{year}{2015}).

\end{thebibliography}
\end{document}